\documentstyle[prl,aps,twocolumn,epsfig]{revtex}

\begin{document}
\draft

\title{Spin blockade in ground state resonance of a quantum dot}

\author{A.~K.~H\"uttel,$^1$ H.~Qin,$^1$ A.~W.~Holleitner,$^1$ 
R.~H.~Blick,$^{1,\ast}$ K.~Neumaier,$^2$\\
D.~Weinmann,$^3$ K.~Eberl,$^4$ and 
J.~P.~Kotthaus$^1$}

\address{$^1$ Center for NanoScience and Sektion Physik, 
Ludwig--Maximilians--Universit\"at, Geschwister--Scholl--Platz 1, 
80539~M\"unchen, Germany.}

\address{$^2$ Walther-Meissner-Institut f\"ur Tieftemperaturforschung,
Walther-Meissner-Stra{\ss}e 8, 85748 Garching, Germany.}

\address{$^3$ Institut de Physique et Chimie des Mat\'{e}riaux de
  Strasbourg, UMR 7504 (CNRS-ULP), 23 rue du Loess,
  67037 Strasbourg, France.}

\address{$^4$ Max-Planck-Institut f\"ur Festk\"orperforschung, 
Heisenbergstra{\ss}e 1, 70569 Stuttgart, Germany.}

\date{\today{ }}
\maketitle
\begin{abstract}
We present measurements on spin blockade in a laterally integrated
quantum dot. The dot is tuned into the regime of strong Coulomb
blockade, confining $\sim 50$ electrons. At certain electronic states 
we find an additional mechanism suppressing electron transport. This we 
identify as spin blockade at zero bias, possibly
accompanied by a change in orbital momentum in subsequent dot ground states. 
We support this by probing the bias, magnetic field
and temperature dependence of the transport spectrum. Weak violation of the blockade is
modelled by detailed calculations of 
non-linear transport taking into account forbidden transitions. 
\end{abstract}

Conventional electronics relies on controlling charge in semiconductor
transistors. The ultimate 
limit of integration is reached when these transistors,
termed quantum dots, are operated by exchanging single electrons only.
The mechanism governing
electron transport through dots is known as Coulomb blockade
(CB)~\cite{ashoori96:413,kouwenhoven_review}. 
Apart from this, electrons naturally possess a spin degree of freedom, which  
recently attracted considerable interest regarding the combination of
spintronics and quantum information 
processing~\cite{spintronics}. Hence, studies on the interplay of spin and
charge quantum states in quantum dots form an integral
contribution for defining and controlling electron spin quantum bits. 

One of the key techniques applied to study electronic structure in quantum
dots is transport spectroscopy: Defining the dots by locally depleting a
two-dimensional electron gas enables this direct monitoring
of ground and excited states of the artificial atoms~\cite{johnson}.
In contrast to Kondo physics~\cite{kondo} in the limit of
transparent tunneling barriers between the quantum dot and the reservoirs, we are
focusing on the regime of opaque barriers and Coulomb blockade. In other words
the strong hybridization with electronic reservoir
states found for the dot electrons in Kondo physics
is suppressed and both electron spin and orbital quantum numbers remain well-defined.

Inspired by an early experiment of Weis~{\it et al.}~\cite{weis}
it was suggested by Weinmann~{\it et al.}~\cite{weinmann} that
transport through single quantum dots can be blocked due to spin effects.
Spin selection rules particularly prohibit single
electron tunneling (SET) transitions between $N$ and $N+1$ electron ground 
states of the dot which differ in total spin by $\Delta S > 1/2$.  
This phenomenon was termed spin blockade (SB) (type-II), occuring in addition to 
conventional CB and leading to a
suppression of the corresponding conductance peak in linear transport~\cite{weinmann}. 

Here, we demonstrate a spin blockade effect in the many-electron ($N\sim 50$)
limit in a two-dimensional quantum dot, in contrast to earlier measurements by Rokhinson
{\it et al.}~\cite{rok} on a small three-dimensional silicon quantum dot with only a few
electrons.
In particular we present detailed measurements on the bias and magnetic field 
dependence of the transport spectrum of our laterally gated quantum dot.  
Weak violation of spin blockade due to spin orbit coupling is found. Modelling the system
by numerical calculations taking into account type-II spin blockade and weak
spin relaxation, we observe excellent agreement with the transport spectrum.

\begin{figure}[tb]
\begin{center}
\epsfig{file=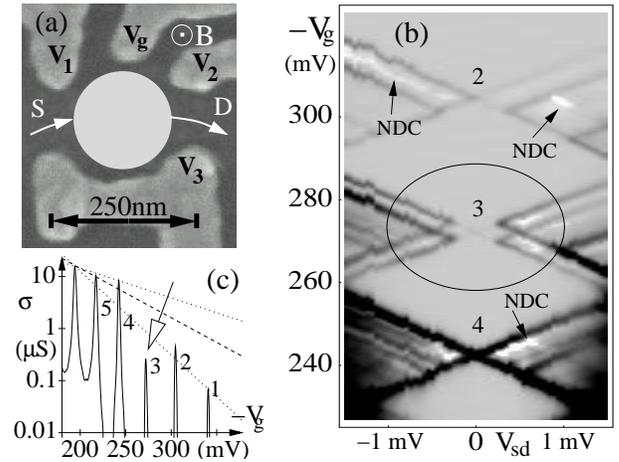, width=8cm}
\end{center}
\caption{
(a) SEM micrograph of the gate electrodes defining the quantum
dot (top view). Approximate dot area (circle), source (S)
and drain (D) contacts are marked schematically. 
(b) Conductance measurement on the quantum dot ($B=0\,\mathrm{T}$; white denotes
$-0.1\,\mu\mathrm{S}$ (NDC),  
grey (large areas) $0\,\mu\mathrm{S}$, 
black $\ge 2.0\,\mu\mathrm{S}$).
(c) Conductance trace at $V_{\mathrm sd}=0\,$V; an exponential fit
through the conductance peak maxima has been added (dashed line).
Conductance peak three is suppressed by spin blockade.
}
\label{overview}
\end{figure}

The quantum dot we use is shown in Fig.~\ref{overview}:
A number of split gates is defined by
electron beam lithography on top of an AlGaAs/GaAs
heterostructure.
When applying negative gate voltages $V_1$, $V_2$, $V_3$, and $V_g$, a single nearly circular 
quantum dot is formed in the
two-dimensional electron system (2DES) 
$120\,\mathrm{nm}$ below the surface. 
At $4.2\,\mathrm{K}$ the carrier density of the 2DES is $n_s = 1.8
\times
10^{15}\,\mathrm{m}^{-2}$ and 
the electron mobility is $75\,\mathrm{m}^{2}/\mathrm{Vs}$. 
For the measurements presented here a dot with an electronic
diameter of approximately $90\,\mathrm{nm}$ was defined. 
The 2DES was cooled to a bath
temperature of $23\,\mathrm{mK}$ 
and an electron temperature of
$T_{\mathrm{el}}=95\,\mathrm{mK}$
in a $^{3}$He/$^{4}$He dilution
refrigerator. 
Using an excitation voltage of $12\,\mu\mathrm{V}$ at
$18\,\mathrm{Hz}$,
the noise
in the CB regime is lower than $50\,\mathrm{nS}$. 
Electronic radius and mean level spacing both lead to an estimate of $N \sim
50$ electrons on the dot.

In Fig.~\ref{overview}(b) the conductance is plotted as a function of gate voltage
$V_g$  and source/drain bias $V_{\mathrm{sd}}$. 
Near $V_g=-290\,\mathrm{mV}$, a capacitance ratio of $\alpha=C_g/C_\Sigma=0.059$ has been
obtained from this
measurement, leading to a total dot capacitance of
$C_\Sigma=83\,\mathrm{aF}$.  
Therefore the Coulomb
charging energy $E_C = e^2 / 2 C_{\Sigma}$ of around
$1.9\,\mathrm{meV}$, giving the electrostatic
energy required to add an electron to the quantum dot, is dominant
compared to the spatial
quantization energy
$\epsilon$.  Nevertheless, a rich spectrum of excited state resonances
in all SET 
regions is revealed. In addition, multiple lines of negative differential
conductance (NDC) are visible, as
predicted in Ref.~\cite{weinmann}. Hence, we can conlude that in these
regions spin polarized excited states lead to a reduction of current via spin
blockade type-I. 

As a striking feature, the conductance at peak three in Fig.~\ref{overview}
is, for small $\left|V_{\mathrm{sd}}\right|$, 
significantly below the expected value as compared to
peaks two and four. 
Only for $\left|V_{\mathrm{sd}}\right| \ge 300\,\mathrm{mV}$
transport via an excited state becomes possible, and the conductance
increases. Such strong suppression of ground state tunneling
is caused by spin blockade of type-II.
This phenomenon involves ground state transitions only.
An electron is blocked from entering the dot,
since the transition involves two levels with total spin difference
$\Delta S > 1/2$. 

Fig.~\ref{overview}(c) shows the conductance trace at
$V_{\mathrm{sd}} = 0$ in logarithmic scale.
As the width of the tunneling barriers increases at higher
$\left|V_g\right|$, the maximum current
approaches zero in the case of complete pinch-off. 
In a simple quantum-mechanical picture the maximum peak conductance is
assumed to decrease exponentially with $-V_{\rm g}$. Again, the amplitude of peak
three is considerably below the most general maximum/minimum bounds (dotted
lines in Fig.~\ref{overview}(c)).
An exponential fit through the peak maxima (dashed line) allows us to estimate a
lifetime of the spin state causing spin blockade.
We assume that the tunneling rates $\Gamma_{L/R}/h$ of the 
left and right barrier defining the dot
are equal -- which appears reasonable because of the
high degree of symmetry shown in the diamond conductance
structure of Fig.~\ref{overview}(b).

In our case of weak coupling to the reservoirs, where
$\Gamma \ll k_B T \ll \epsilon < E_C$, the maximum conductance through the
quantum dot is given by~\cite{beenakker} 
$$
\sigma_{\mathrm{max}}=\frac{e^2}{h} \frac{1}{4k_B T_{\mathrm{el}}} \frac{\Gamma}{2}. 
$$
The dwell time for an electron in the system is estimated with  
$\tau_{b}=h/\Gamma_{b}=9.5\,\mathrm{ns}$ in the case of 
transport blockade compared to $\tau_{e}=h/\Gamma_{e}=0.8\,\mathrm{ns}$ as
extrapolated value without any spin effects~\cite{yacoby}.
Therefore, coupling of the long-lived state to its environment corresponds
to a time scale of $\Delta\tau= \tau_{b}-\tau_{e} \approx 8.7\,\mathrm{ns}$, 
i.e. the high-spin state survives for several nanoseconds.

\begin{figure}[tb]
\begin{center}
\epsfig{file=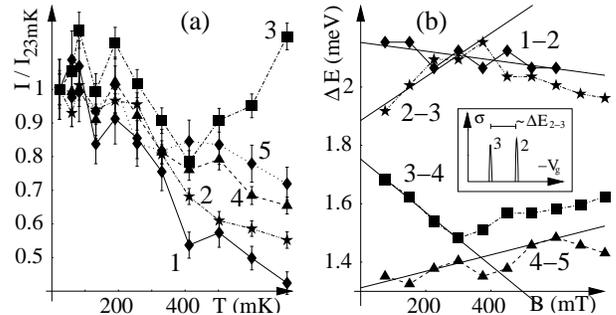, width=8cm}
\end{center}
\caption{
(a)
Relative temperature dependence of the CB peak height (dc measurement
at $V_{\mathrm{sd}}=20\,\mu$V); only peak three increases with temperature above $T_e
\sim 400\,\mathrm{mK}$, in every
other case the decay characteristic for CB ist apparent. 
(b)
$B$-dependence of the single electron addition energy $\Delta E_{N-(N+1)} =
\alpha(V_g) \left| V_{g,N} 
- V_{g,(N+1)}\right| $. For small $B$, peak spacings around peak three vary
rapidly with respect to a magnetic field perpendicular to the 2DES.
        }
\label{further}
\end{figure}

Additional information on spin blockade is given by the
temperature dependence of the current peaks at the SET
resonances, taken from a measurement at $V_{\mathrm{sd}}=20\,\mu\mathrm{V}$ (dc)
and displayed in
Fig.~\ref{further}(a). In the present case of CB ($E_C >
\epsilon \gg k_B T$), a decrease of the current at higher temperatures
is expected~\cite{beenakker}, while peak three shows an increase above
$T_e \sim 400\,\mathrm{mK}$. This gives rise to the assumption that
relaxation becomes accessible at an energy scale
comparable to $k_B T_e \approx 35\,\mu eV$~\cite{weinmann}, by order of
magnitude consistent with
a change in spin configuration at $B=0\,\mathrm{T}$ \cite{koskinen}.

In Fig.~\ref{further}(b), we observe the change in electron addition energy $\Delta E =
E_C + \Delta \epsilon$
with increasing magnetic field, which is proportional to
the gate voltage SET peak spacing (see inset of Fig.~\ref{further}(b)). 
Whereas a magnetic field parallel to the 2DES couples primarily to the
electron spin via the Zeeman energy term, the perpendicular field applied here
strongly influences the orbital states as well. 
Most observed peaks display a weak field dependence. Strikingly, the
spin-blocked resonance is shifting strongly at low magnetic fields, as can bee
seen at hand of the high slopes of charging energy 2-3 and 3-4 in Fig.~\ref{further}(b).

Different explanations for this phenomenon are possible. On one hand, 
recent measurements by Pallecchi {\it et al.} \cite{pallecchi} 
indicate the possibility of greatly enhanced $g$-factors in 2D
quantum dots; values of up to $g\sim 20$ have been observed in AlGaAs/GaAs
quantum wires. As an example, at $g\approx 5.4$ line shifts
can be approximated solely by the Zeeman shift caused by a spin
difference $\Delta S=5/2$ of subsequent electron number ground states
(cf. Ref. \cite{altshuler}), thus explaining spin blockade. At nearby,
non-blocked resonances, the peak shift 
is then consistent with a spin change in the dot of $\Delta S=1/2$, i.e. the
addition of a single electron spin.
Theoretical predictions \cite{matveev} support a $g$-factor deviation in
case of unusually large spin-orbit interaction. 

However, orbital states are modified by the magnetic field
perpendicular to the 2DES as well. Hence, on the other hand a change in 
orbital quantum numbers, particularly
ground state angular momentum, is an alternative explanation for the peak shift. 
Since momentum and angular momentum of an electron are not
preserved when traversing the quantum point contacts~\cite{reed}, 
a misalignment of spatial quantum numbers
alone will not lead to a total blocking of transport. Therefore, a change in both $L$ 
and $S$ between subsequent electron number ground states leads to a scenario
explaining the data.

\begin{figure}[tb]
\begin{center}
\epsfig{file=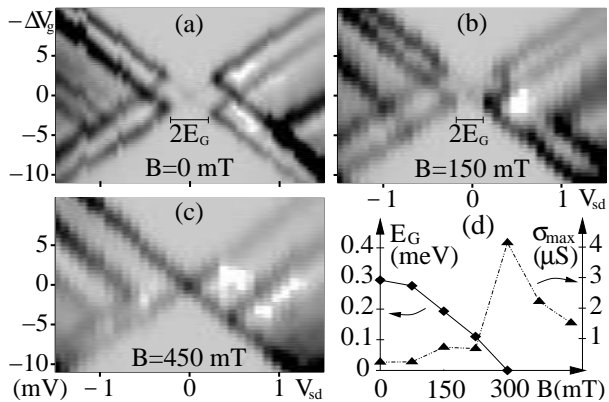, width=8cm}
\end{center}
\caption{
(a) to (c): Effect of an external perpendicular
magnetic field on the spin blocked SET resonance three. 
(a) Detail from Fig.~\ref{overview}(b). (b) At $B = 150\,\mathrm{mT}$ the gap
has partly closed. 
Conductance is still suppressed
for $|V_{\mathrm{sd}}| <  200\,\mu\mathrm{V}$.
(c) Blockade phenomena 
disappear completely at $450\,\mathrm{mT}$. Simultaneously regions of
NDC are evolving. 
(d) Magnetic field dependence of the transport blockade gap as indicated in
(a) and the maximum conductance at $V_{\mathrm{sd}}=0\,\mathrm{V}$. At
approximately $300\,\mathrm{mT}$ the gap is quenched.
        }
\label{bfield}
\end{figure}%

In addition nonlinear conductance measurements at finite magnetic field
perpendicular to the 2DES have
been taken. In Fig.~\ref{bfield}(a) to (c) the transport-blocked SET resonance is
shown at $0\,\mathrm{mT}$, $150\,\mathrm{mT}$, and
$450\,\mathrm{mT}$. At a field strength of only $300\,\mathrm{mT}$ the
quantum levels in the dot are already shifted sufficiently to reenable ground state
transport. This is also demonstrated in Fig.~\ref{bfield}(d), which shows 
the conductance at $V_{\mathrm{sd}}=0\,\mathrm{V}$ and the
blockade gap energy $E_G$ as function of $B$. At $B=300\,\mathrm{mT}$ a strong 
increase in conductance is seen. Here, one quantum ground state participating in SET
changes because of a level crossing; for higher $B$, 
$\left| \Delta S \right| = 1/2$ for the $N$ and $N+1$ electron ground state and SET
transport takes place. This assumption is directly supported by the data of
Fig.~\ref{further}(b), $B=300\,\mathrm{mT}$ being the field strength where the
$B$-dependence of addition energies around peak three adapts to the one around nearby peaks.
For high $B$ the overall conductance through the quantum dot is decreasing
because of a gradual compression of the dot states by the magnetic field~\cite{datta}.

The data indicate strong electronic
correlations in the quantum dot involving both spin and orbit of the wavefunction.
The expected impact of correlation effects can be estimated by the 
conventional parameter $r_s= 1/\sqrt{\pi n_s} 
a_B$  \cite{chui} with $a_B$ as the Bohr radius in GaAs.
In our sample, it is given by $r_s=1.2$,
thus even assuming a somewhat lower electron density in the dot,
it remains still far below critical values of $r_s \sim 8$ to
$36$. Taking into account the high number of
electrons present ($N\approx 50$), the measurement presents
an unusually strong deviation from
the constant interaction model~\cite{kouwenhoven_review}.

\begin{figure}[tb]
\begin{center}
\epsfig{file=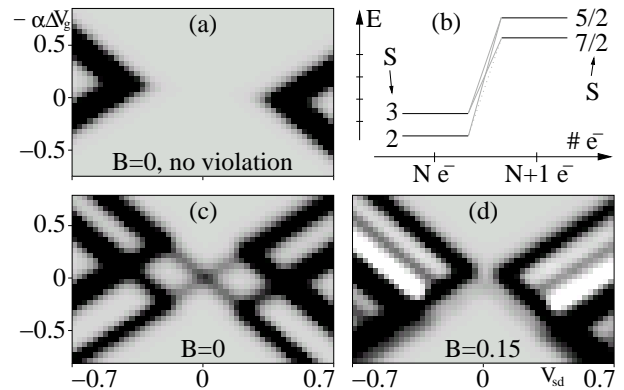, width=8cm}
\end{center}
\caption{
Model calculation of nonlinear transport for spin blockade
type-II. (a) Transport spectrum in the case of perfect spin
conservation, differing considerably from measured data. (b) Schematic
representation of the quantum levels
assumed for the calculation. (c) When taking into account a weak violation of
spin transition rules, 
the characteristic blockade situation at $B=0\,\mathrm{T}$ (cf. 
Fig.~\ref{overview}(b)) emerges. In (d) spin blockade is partially lifted by a magnetic field,
cf. Fig.\ref{bfield}. 
        }
\label{calc}
\end{figure}%

A spin blockade level scenario and results from
numerical transport calculations are
depicted in Fig.~\ref{calc}. The calculations were performed using a
master equation approach, describing the regime of sequential 
tunneling, as in Ref.~\cite{weinmann}. The transition rates between
the many-body states of the dot include a Clebsch-Gordon factor which
accounts for the spin selection rules. Spin values of $S=2$ and $3$ have
been assumed in the model for the ground and excited 
$N$-electron states, $7/2$ and $5/2$ for the $(N+1)$-electron states, 
respectively. If one assumes perfect spin conservation in tunneling  
processes and the total spin of the electrons confined in the 
dot to be a good quantum number, the theory yields transport spectra
which are qualitatively different from the experimental ones (see Fig.~\ref{calc}(a)).
This comparison of the theory and the measurements indicates that 
spin blockade is weakly violated, possibly
due to spin-orbit coupling. Such a state mixing mechanism in the
quantum dot leads to a violation of both spin and angular momentum
conservation. The couplings might differ considerably in our case from
previous observations~\cite{rok}; possible
reasons include the material system (AlGaAs/GaAs instead of Si), 
the potential shape of the dot, and their effect on the details 
of the spin-orbit Hamiltonian.

We take this effect into account by including a phenomenological 
spin-independent transition rate between many-body states in our model, 
allowing for electron tunneling processes from and into the dot 
in which the spin selection rules are violated. Using the small
value of 1\% of the spin-allowed transition rates, and assuming
quantum levels as displayed in Fig.~\ref{calc}(b), excellent 
agreement of theory and experiment is found in Fig.~\ref{calc}(c) and (d). 
The magnetic field was introduced as a Zeeman-like shift of
the levels, leading to a level crossing in the $N$-electron spectrum
at $B/g\mu_B \approx 0.17$. This explains that the spin blockade is
lifted at stronger magnetic field, in agreement
with the experimental observations.

In conclusion, we have demonstrated transport blockade in 
a quantum dot containing approximately 50
electrons, caused by spin selection rules: electronic correlations lead to a
discontinuity in both ground 
state spin and spatial quantum numbers for subsequent electron
counts. As a result, ground state transport blockade caused by spin
selection rules is found.

We observe the properties predicted
for such a system: Spin blockade can be lifted by raising the 
temperature, or applying a source/drain voltage. Shifting the quantum
levels via a magnetic field leads to a change in both spatial and spin quantum
numbers, lifting spin blockade as well. Two alternative mechanisms for the
observed level shift are proposed, firstly, a large Zeeman shift via
$g$-factor enhancement, secondly, a combined change of spin and orbital
quantum numbers.

In the measurement, spin blockade is weakly violated. A likely candidate 
for this conduction process is given by the spin-orbit interaction induced spin state mixing 
in the quantum dot, being consistent with both proposed mechanisms of level shifting. 
Taking a corresponding weak violation of selection rules into account, numerical
calculations of a straightforward spin blockade model lead to an excellent
agreement of theory and experiment, and the properties of the nonlinear
transport spectrum at zero and large $B$ are accurately reproduced.

We like to thank L.~Rokhinson for detailed discussions.
We acknowledge financial support by the Deutsche
Forschungs\-ge\-mein\-schaft
(Bl/487-2-2, SFB-348) and the Bundesministerium f\"ur
Forschung und Technologie (project 01BM/914). Many thanks to the
Stu\-dien\-stif\-tung des deutschen Volkes
and the Stif\-tung Ma\-xi\-mi\-liane\-um (AKH), the
Volkswagenstiftung (HQ), and the European Union (DW, RTN program) for
their support.

$^{\ast}$Corresponding author - e-mail:
robert.blick@physik.uni-muenchen.de 


\bibliographystyle{prsty}

\end{document}